\documentclass[11pt,a4paper]{article}
\def\bea{\begin{eqnarray}}
\def\eea{\end{eqnarray}}

\def\beq{\begin{equation}}
\def\eeq{\end{equation}}
\def\ba{\beq\new\begin{array}{c}}
\def\ea{\end{array}\eeq}
\def\be{\ba}
\def\ee{\ea}

\makeatletter
\newdimen\normalarrayskip              
\newdimen\minarrayskip                 
\normalarrayskip\baselineskip
\minarrayskip\jot
\newif\ifold             \oldtrue            \def\new{\oldfalse}
\def\arraymode{\ifold\relax\else\displaystyle\fi} 
\def\eqnumphantom{\phantom{(\theequation)}}     
\def\@arrayskip{\ifold\baselineskip\z@\lineskip\z@
     \else
     \baselineskip\minarrayskip\lineskip2\minarrayskip\fi}
\def\@arrayclassz{\ifcase \@lastchclass \@acolampacol \or
\@ampacol \or \or \or \@addamp \or
   \@acolampacol \or \@firstampfalse \@acol \fi
\edef\@preamble{\@preamble
  \ifcase \@chnum
     \hfil$\relax\arraymode\@sharp$\hfil
     \or $\relax\arraymode\@sharp$\hfil
     \or \hfil$\relax\arraymode\@sharp$\fi}}
\def\@array[#1]#2{\setbox\@arstrutbox=\hbox{\vrule
     height\arraystretch \ht\strutbox
     depth\arraystretch \dp\strutbox
     width\z@}\@mkpream{#2}\edef\@preamble{\halign
\noexpand\@halignto
\bgroup \tabskip\z@ \@arstrut \@preamble \tabskip\z@ \cr}%
\let\@startpbox\@@startpbox \let\@endpbox\@@endpbox
  \if #1t\vtop \else \if#1b\vbox \else \vcenter \fi\fi
  \bgroup \let\par\relax
  \let\@sharp##\let\protect\relax
  \@arrayskip\@preamble}
\def\eqnarray{\stepcounter{equation}%
              \let\@currentlabel=\theequation
              \global\@eqnswtrue
              \global\@eqcnt\z@
              \tabskip\@centering
              \let\\=\@eqncr
              $$%
 \halign to \displaywidth\bgroup
    \eqnumphantom\@eqnsel\hskip\@centering
    $\displaystyle \tabskip\z@ {##}$%
    \global\@eqcnt\@ne \hskip 2\arraycolsep
         $\displaystyle\arraymode{##}$\hfil
    \global\@eqcnt\tw@ \hskip 2\arraycolsep
         $\displaystyle\tabskip\z@{##}$\hfil
         \tabskip\@centering
    &{##}\tabskip\z@\cr}
\begingroup\ifx\undefined\newsymbol \else\def\input#1 {\endgroup}\fi
\input amssym.def \relax
\input amssym
\newfont{\hr}{msbm10}
\newfont{\ams}{msam10}

\font\numbers=cmss12
\font\upright=cmu10 scaled\magstep1
\def\stroke{\vrule height8pt width0.4pt depth-0.1pt}
\def\topfleck{\vrule height8pt width0.5pt depth-5.9pt}
\def\botfleck{\vrule height2pt width0.5pt depth0.1pt}
\def\Zmath{\vcenter{\hbox{\numbers\rlap{\rlap{Z}\kern 0.8pt\topfleck}\kern
2.2pt
                   \rlap Z\kern 6pt\botfleck\kern 1pt}}}
\def\Qmath{\vcenter{\hbox{\upright\rlap{\rlap{Q}\kern
                   3.8pt\stroke}\phantom{Q}}}}
\def\Nmath{\vcenter{\hbox{\upright\rlap{I}\kern 1.7pt N}}}
\def\Cmath{\vcenter{\hbox{\upright\rlap{\rlap{C}\kern
                   3.8pt\stroke}\phantom{C}}}}
\def\Rmath{\vcenter{\hbox{\upright\rlap{I}\kern 1.7pt R}}}
\def\Z{\ifmmode\Zmath\else$\Zmath$\fi}
\def\Q{\ifmmode\Qmath\else$\Qmath$\fi}
\def\N{\ifmmode\Nmath\else$\Nmath$\fi}
\def\C{\ifmmode\Cmath\else$\Cmath$\fi}
\def\R{\ifmmode\Rmath\else$\Rmath$\fi}

\newcounter{app}

\def\app{\setcounter{equation}{0}
\def\theequation{\Alph{app}.\arabic{equation}}\par
   \addvspace{4ex}
   \@afterindentfalse
  \secdef\@app\@dapp}

\newcommand\@app{\@startsection {app}{1}{0ex}%
                                   {-3.5ex \@plus -1ex \@minus -.2ex}%
                                   {2.3ex \@plus.2ex}%
                                   {\normalfont\Large\bf}}
\def\@dapp#1{%
{\parindent \z@ \raggedright  \bf #1}\par\nobreak}
\def\l@app#1#2{\ifnum \c@tocdepth >\z@
    \addpenalty\@secpenalty
    \addvspace{1.0em \@plus\p@}%
    \setlength\@tempdima{8em}%
    \begingroup
      \parindent \z@ \rightskip \@pnumwidth
      \parfillskip -\@pnumwidth
      \leavevmode \bfseries
      \advance\leftskip\@tempdima
      \hskip -\leftskip
      #1\nobreak\hfil \nobreak\hb@xt@\@pnumwidth{\hss #2}\par
    \endgroup\fi}
\newcounter{sapp}[app]

\def\sapp{\def\theequation{\Alph{app}.\arabic{equation}}
\par
\@afterindentfalse
  \secdef\@sapp\@dsapp}
\newcommand{\@sapp}{\@startsection{sapp}{2}{\z@}%
                                     {-3.25ex\@plus -1ex \@minus -.2ex}%
                                     {1.5ex \@plus .2ex}%
                                     {\normalfont\large\bfseries}}

\def\@dsapp#1{%
{\parindent \z@ \raggedright  \bf #1
}\par\nobreak}
\newcommand{\l@sapp}{\@dottedtocline{2}{1.5em}{2.3em}}

\def\2{{1\over 2}}
\def\N2{${\cal N}=2$}

\def\be{ \begin{eqnarray} }
\def\ee{ \end{eqnarray} }

\def\bea{\begin{eqnarray}}
\def\eea{\end{eqnarray}}

\def\beq{\begin{equation}}
\def\eeq{\end{equation}}
\def\ba{\beq\new\begin{array}{c}}
\def\ea{\end{array}\eeq}
\def\be{\ba}
\def\ee{\ea}

\def\cn{\hbox{cn}}
\def\sn{\hbox{sn}}

\title{Self-dual Hamiltonians as Deformations of Free Systems
}

\author{A.Mironov\thanks{mironov@lpi.ac.ru, mironov@itep.ru} \
\\ \normalsize \em  Theory
Department, Lebedev Physics Institute, Moscow
~117924, Russia
\\
and
\\
\normalsize \em ITEP, Moscow
117259, Russia }

\date{}

\begin{document}

\renewcommand{\thepage}{}

\maketitle

\vspace{-6.7cm}

\begin{center}
\hfill FIAN/TD-08/01\\
\hfill ITEP/TH-18/01\\
\hfill hep-th/0104253
\end{center}

\vspace{4cm}

\begin{abstract}
We formulate the problem of finding self-dual Hamiltonians (associated
with integrable systems) as deformations of free systems given on various
symplectic manifolds and discuss
several known explicit examples, including recently found double elliptic
Hamiltonians. We consider as basic the notion of self-duality, while the
duality in integrable systems (of the Toda/Calogero/Ruijsenaars type)
comes as a derivative notion (degenerations of
self-dual systems).

This is a talk presented at the Workshop ``Classical and Quantum Integrable
Systems", Protvino, January, 2001.
\end{abstract}

\paragraph{1.}

The new energy that theory of integrable many-systems obtained during last
several years is mostly due to the unexpected discovery of relations between
the $N=2$ supersymmetric gauge theories \cite{SW} and integrable systems
\cite{GKMMM} (see also \cite{SWbook} and references therein). In fact, this
latter still gets new motivations from the physical base.

For instance, the needs in description of $6d$ gauge theories have led to the
notion of double-elliptic systems \cite{BMMM3,FGNR}. These systems have been
constructed using the notion of duality \cite{BMMM3}, which is also obliged
to a physical set-up \cite{GNR}.

In fact, the duality in integrable systems was first observed by
S.Ruijsenaars \cite{R} in rational and trigonometric Calogero and Ruijsenaars
many-body systems. Since then, it was extended to the elliptic systems
\cite{BMMM3,FGNR,dell}. However, the very notion of duality was not
unambiguously formulated so far. Moreover, it looks so that there are plenty
of dual systems, constructing them does not look a problem at all!

In this short note we discuss that one can start from a more restrictive
notion of self-duality. It can be easier defined, naturally leads
to duality and, being more restrictive, immediately put a problem of
finding more/all self-dual systems.
Here we restrict ourselves with just several
manifest examples of self-dual systems.  We also do not discuss the
symplectic geometry behind the notion of self-duality.

The standard facts and known examples of dual systems, as well as a proper
technique to deal with them can be found in reviews \cite{revdual}.

\paragraph{2.}

Let us consider a free Hamiltonian system with $N$ degrees of freedom
whose phase space (symplectic manifold) is just $\R^{2N}$. In Darboux
coordinates the symplectic form is $\sum_i dp_i\wedge dq_i$. Then, there are
$N$ independent functions on the phase space, integrals of motion in
involution whichever one of them chosen to be the Hamiltonian of the system.
The choice of these integrals is quite free, say, they can be just symmetric
powers of $p_i$:  $H_k(p,q)=\sum_i p_i^k$.

There are no more
independent integrals in involution, say, similar functions of
coordinates. Therefore, this choice in a sense breaks the symmetry between
permutations of $\{p\}$ and $\{q\}$ (i.e. fixes the polarization). In order
to restore this symmetry, one should perform an (anti)canonical transformation
$p_i\to Q_i$, $q_i\to P_i$. In the new coordinates
$\tilde H_k(P,Q)=H_k(p,q)=\sum_i Q_i^k$, while $\sum_i q_i^k=H_k(P,Q)$. Here
$\tilde H_k(P,Q)$ are integrals (Hamiltonians) after the canonical
transformation is done.

This means that the action of the symplectic group on
the Hamiltonians should be provided with a proper (canonical) change of
variables on the symplectic manifold (a kind of covariance property). It can
be also given by the pair of relations

\be\label{sdf}
H_k(p,q)=\sum_i Q_i^k\\
H_k(P,Q)=\sum_i q_i^k
\ee
These relations manifestly realize the symmetry between $\{Q\}$ (essentially,
old $\{p\}$) and $\{q\}$.

Now let us consider a less trivial, interacting system. We ask it to be integrable
in the Liouville sense, i.e. still to possess $N$ independent integrals of
motion in involution. Then, the
Hamiltonians can be quite tricky functions on the phase space. One may ask
how to generalize the symmetry between coordinates and momenta to this case.
In order to do this, one can just use the same relations (\ref{sdf}).

This
means in interacting system that, at the first stage, one makes a canonical
change to the free system (essentially, action variables) and then makes the
anticanonical transformation just permuting coordinates and momenta. Note
that the relations (\ref{sdf}) already contains $2N$ equations, and there is
additionally the condition of canonicity of the transformation
$p_i\to Q_i$, $q_i\to P_i$. This restricts the Hamiltonians.

Now let us consider a more general situation when the phase space is not
flat and may be compact. Then, even a free system possesses less trivial
Hamiltonians $h_k(p)$, since they are to be good functions on the phase
space.  Then, the system (\ref{sdf}) should be substituted by

\be\label{sd}
H_k(p,q)=h_k(Q)\\
H_k(P,Q)=h_k(q)
\ee

{\large \bf Definition.}  The system with Hamiltonians that satisfy (\ref{sd}) with the variables
$(p,q)$ and $(P,Q)$ related by a canonical transformation we call
{\bf self-dual}.

We do not know any constructive way to build such systems, and,
therefore, just restrict ourselves here with several explicit examples.

Note that (\ref{sd}) enjoys a kind of covariance property: instead of a set
$h_k$ of free Hamiltonians, one can use any other set of proper functions on
the phase space. All of them are some functions of the originally fixed
$h_k$, while the new dual Hamiltonians are the same functions of the old
ones.

\paragraph{3.}

We shall consider the complexified version of the phase space so that
Hamiltonians should celebrate good analytic properties on the phase space.
However, the typical situation is when the phase space is very asymmetric
under the change of polarization, i.e. under interchanging the coordinates
and momenta.  In this case one can not use the equations (\ref{sd}).

Our strategy below in such a case will be to construct a self-dual system on
the non-trivial phase space that {\it depends on moduli} and then to go to
the boundary of the moduli space producing different degenerate situations,
where say, only $q$ but not $Q$ gets into degeneration region. In the vicinity
of the boundary the two lines in (\ref{sd}) look
different

\be\label{d}
H_k^{(1)}(p,q)=h_k^{(1)}(Q)\\
H_k^{(2)}(P,Q)=h_k^{(2)}(q)
\ee
and give rise to two {\it different} sets of Hamiltonians. We call
them {\bf dual}. The dual Hamiltonians are just the Hamiltonian of the
self-dual system taken in {\it two different} regions of the moduli space. In
other words, this means that one should use just the same equations
(\ref{sd}) but in their two lines properly interchange points of the moduli
space: $H_k^{(1,2)}(p,q)=H_k(p,q;m_{1,2})$, $h_k^{(1,2)}(q)=h_k(q;m_{1,2})$.

Therefore, one can get the counterpart of self-duality for more
complicated phase-spaces via degeneration. In fact, the inverse is
also correct:  the dual system proved to be a powerful technical tool for
constructing the self-dual Hamiltonians to start with the dual ones, since
these latter are often constructed simpler.

Let us start with a system with one degree of freedom (it also includes
systems with two degrees of freedom with decoupling the centre of mass).
In this case, the new variables $P$ and $Q$ are given by the two
relations (\ref{sd}) and one should impose the requirement of canonicity
of the transformation:

\be\label{can}
{\partial H(p,q)\over\partial p}h'(q)={\partial H(P,Q)\over P}h'(Q)
\ee
Using again {\ref{sd}) one now rewrites the r.h.s. of (\ref{can})
as a function of $p$ and $q$ and obtains a complicated
functional differential equation that defines the Hamiltonian of the
self-dual system. Certainly, we are not able to solve this equation and
even to say how many solutions it has. However, this is one equation for
one function.

In a system with $N$ degrees of freedom, the relation (\ref{sd}) again
defines the new coordinates and momenta, while the canonicity condition
gives a system of equations for the Hamiltonian.

\paragraph{4.}

Now we turn to the concrete examples. We start with the simplest case of the
system with one degree of freedom. The free Hamiltonians below are chosen
in the form that leads to some standard Hamiltonian systems.

\begin{itemize}
\item

The phase space ${\cal M}=\R^2$. The free system is given by the Hamiltonian
$H=p^2/2$. A solution to the self-duality equations (\ref{sdf}) is
$H=p^2/2+g^2/q^2$, where $g$ is an arbitrary parameter. This is the rational
Calogero Hamiltonian with coupling constant $g$.

\item

The phase space is ${\cal M}=S^1\times S^1$, the product of two circles of
radii $R_1$ and $R_2$. The free system is given by the Hamiltonian $\cos
(p/R_1)$.  A self-dual Hamiltonian is  the Ruijsenaars Hamiltonian
$H=\sqrt{1-2g^2/\sin^2 (q/R_2)}\cos
(p/R_1)$. Here the radii of the circles
give a moduli space of the theory. The self-duality relation is

\be\label{sdt}
H(p,q;R_1,R_2)=\cos(Q/R_1)\\
H(p,q;R_2,R_1)=\cos (q/R_2)
\ee

By degenerating the momentum circle ($R_1\to\infty$), one gets the
trigonometric Calogero Hamiltonian $H_{tC}$, while degenerating the
coordinate one, one obtains the rational Ruijsenaars Hamiltonian $H_{rR}$ so
that the relation (\ref{sdt}) becomes the duality relation:

\be\label{dt}
\begin{array}{rl}
(H(p,q;R_1=\infty,R_2)\equiv )& H_{tC}(p,q)=Q^2/2\\
(H(p,q;R_2,R_1=\infty)\equiv )& H_{rR}(P,Q)=\cos (q/R_2)
\end{array}
\ee

\item
One can consider the complexified phase space of the previous problem, i.e.
the coordinate and momentum living on a cylinder. Then, there is another
solution to the self-dual equations, that with trigonometric functions
substituted with the hyperbolic ones.

\item
Generalizing the previous item, one can consider the coordinate and momentum
living on torii. The free system is given by the Hamiltonian $\cn (p|k)$,
the elliptic (Jacobi) cosine with the elliptic modulus $k$. In order to
find a solution to the self-dual equations (\ref{sd}) in this case, one can
first solve them in the degenerate case, when, say, momentum torus becomes
sphere and $h^{(1)}(Q)=Q^2/2$ in (\ref{d}) (while $h^{(2)}(q)$ is still $\cn
(q|k)$). Then, one can naturally assume that $H^{(1)}(p,q)$ is the elliptic
Calogero-Moser Hamiltonian, $H(p,q)=p^2/2+g^2/\sn^2 (q|k)$ and
obtain solving (\ref{d})

\be
H(P,Q) = \cn(q|k) = \alpha(Q) \cdot
\cn\left(P\sqrt{k'^2 + k^2\alpha^2(Q)}\ \bigg| \
\frac{k\alpha(Q)}{\sqrt{k'^2 + k^2\alpha^2(Q)}}\right)
\label{dualCal}
\ee
with

\be
\alpha^2(Q) = 1 - \frac{2g^2}{Q^2},
\ \ \ k'^2\equiv 1- k^2
\ee
This result prompts an anzatz for the self-dual case \cite{BMMM3}

\be
H(P,Q|k,\tilde k) = \alpha(Q|k,\tilde k)\cdot
\cn\left(P\;\beta(Q|k,\tilde k)\ |\ \gamma(Q|k,\tilde k)\right)
\ee
Inserting it into the self-dual equations (\ref{sd}) (with $k$ and
$\tilde k$ interchanged) gives

\be
\alpha^2(q|\tilde k,k) =
1 - \frac{2\nu^2}{\sn^2(q|k)}, \ \ \
\beta^2(q|\tilde k,k) = \tilde k'^2 + \tilde k^2\alpha^2(q|k), \ \ \
\gamma^2(q|\tilde k,k) = \frac{\tilde k^2\alpha^2(q|k)}
{\tilde k'^2 + \tilde k^2\alpha^2(q|k)}
\ee
This system is called double-elliptic system \cite{BMMM3}. Further details
can be found in \cite{revdual}.

\end{itemize}

One can similarly find multiparticle generalizations. Some technique for
doing this was developed in \cite{R,B,BMMM3,dell,revdual}.
However, the solutions found so far are only the multi-particle rational
Calogero system, the multi-particle trigonometric Ruijsenaars system and the
multi-particle double-elliptic system constructed not quite manifestly.

\paragraph{5.}

In conclusion, we introduced the notion of self-duality (duality) and
constructed several explicit examples. We discussed only classical Hamiltonian
systems, the generalization to the quantum case is quite immediate
\cite{R,FGNR}.  Indeed, one just needs to consider a pair of sets of
Shr\"odinger equations

\be
{\widehat H_k}(\partial_x,x)\Psi(x;\lambda)=h_k(\lambda)\Psi(x;\lambda)\\
{\widehat H_k}(\partial_\lambda,\lambda)
\Psi(x;\lambda)=h_k(x)\Psi(x;\lambda)
\ee
where the variable $\lambda$ plays the role of (function of) energy in the
first set of equations and $x$ does in the second one.

There are also two other issues left beyond the scope of this note. First,
note that the notion of duality can be generalized to the deformations
of non-free systems. To this end, one just can allow on the r.h.s. of
(\ref{sd}) to be arbitrary Hamiltonians.

The other important issue missed is the geometrical meaning of the
self-duality. As far as it provides a symmetry between the old coordinates
$q$ and the new ones $Q$, one would expect a manifest symplectic group
invariance on these coordinates. It is really observed in examples, although
the realization does not look simple, see \cite{F}.

The author is grateful to H.W.Braden, A.Gorsky, A.Marshakov,
M.Olshanetsky and T.Takebe for discussions and to A.Morozov for valuable
discussions and reading the manuscript. The work was partially supported by
grants INTAS 00-561, CRDF \#6531, RFBR 01-01-00548.

\end{document}